\documentclass{article}

\usepackage{graphicx} 
\usepackage{epstopdf}  
\usepackage{listings}
\usepackage{booktabs}

\usepackage{float}
\usepackage{amsmath}
\usepackage{enumitem}  
\usepackage{tcolorbox} 
\tcbuselibrary{breakable}
\usepackage{subcaption}
\usepackage{multicol}

\newcommand{\keywords}[1]{\textbf{Keywords:} #1}
 
\title{Building crypto portfolios with agentic AI}

\author{Antonino Castelli, Paolo Giudici, Alessandro Piergallini\\ Department of Economics and Department of Computer Engineering, \\University of Pavia}

\date{}

\begin{document}

\maketitle
\begin{abstract}
The rapid growth of crypto markets has opened new opportunities for investors, but at the same time exposed them to high volatility. To address the challenge of managing dynamic portfolios in such an environment, this paper presents a practical application of a multi-agent system designed to autonomously construct and evaluate crypto-asset allocations. Using data on daily frequencies of the ten most capitalized cryptocurrencies from 2020 to 2025, we compare two automated investment strategies. These are a static equal weighting strategy and a rolling-window optimization strategy, both implemented to maximize the evaluation metrics of the \emph{Modern Portfolio Theory (MPT)}, such as Expected Return, Sharpe and Sortino ratios, while minimizing volatility. 

Each step of the process is handled by dedicated agents, integrated through a collaborative architecture in Crew AI. The results show that the dynamic optimization strategy achieves significantly better performance in terms of risk-adjusted returns, both in-sample and out-of-sample. This highlights the benefits of adaptive techniques in portfolio management, particularly in volatile markets such as cryptocurrency markets. The following methodology proposed also demonstrates how multi-agent systems can provide scalable, auditable, and flexible solutions in financial automation.
\end{abstract}
\clearpage

\section{Introduction}
Cryptocurrencies have grown from a niche phenomenon to a mainstream asset class within little more than a decade, spurred in part by the post-2008 low-rate environment. While they promise exceptional returns, their pronounced volatility, heavy tails and regime shifts complicate portfolio construction.

Traditional \emph{buy-and-hold} rules struggle in such high-frequency, high-variance settings. Recent progress in Multi-Agent Systems (MAS) offers a modular way to automate the investment pipeline: specialized agents ingest, clean and analyze data, construct portfolios and report results collaboratively. \emph{TradingAgents} shows how LLM-powered agents for fundamental, sentiment and technical analysis can outperform baselines in live trading environments (Xiao et al., 2025) \cite{xiao2025tradingagents}.MAS typically rely on Large Language Models (LLMs). 

Advances in eXplainable AI (XAI) underline the need for transparency in financial ML applications, including credit-risk assessment \cite{misheva2021explainable}. LLM agents can also merge qualitative judgment with quantitative signals for asset pricing, as in the AAPM framework proposed by \emph{Cheng \& Chin (2024)} \cite{cheng2024aapm}. Beyond investment, LLMs have simulated human-like economic behavior for scalable virtual experiments \cite{horton2023large}, and MAS have improved anomaly detection and real-time risk management in markets \cite{park2024enhancing}. Despite this growing literature, empirical evidence on crypto-asset portfolios remains limited.

This paper fills that gap by deploying a MAS built with Crew AI to compare a static equal-weight mean-variance strategy and a rolling 30-day Sharpe-maximizing strategy on the ten largest cryptocurrencies (2020-2025). The study quantifies the performance benefits of adaptive allocation while illustrating how MAS deliver scalable, auditable, and flexible workflows for volatile markets. Building on recent advances in multi-agent architectures \cite{masterman2024survey} and LLM-based market simulations \cite{zhang2024stockagent}, we propose and compare two MAS-based crypto-allocation strategies, showing that dynamic rebalancing significantly improves performance.

\section{Data \& MAS Setup}

Our study focuses on a panel of daily closing prices for the ten largest crypto assets between 20 August 2020 and 13 March 2025. 

The dataset is composed of 1667 daily closing prices for the ten largest cryptocurrencies, considering market capitalization, over the period from 2020-08-20 to 2025-03-13. The selected cryptocurrencies are the following: Bitcoin (BTC), Ethereum (ETH), Binance Coin (BNB), Solana (SOL), Ripple (XRP), Cardano (ADA), Dogecoin (DOGE), Avalanche (AVAX), Polkadot (DOT), and Shiba Inu (SHIB). The choice of these assets was made because of their consistent market relevance and liquidity.

The descriptive statistics of the data are reported in Table \ref{Tab1}.

\begin{table}[H]
\centering
\caption{Descriptive statistics of selected cryptocurrencies (USD)}
\resizebox{\textwidth}{!}{
\begin{tabular}{lccccc}
\toprule
\textbf{Asset} & \textbf{Mean} & \textbf{Median} & \textbf{Standard Deviation} & \textbf{Min} & \textbf{Max} \\
\midrule
BTC  & 45249.89 & 41770.30 & 21685.75 & 15787.28 & 106146.27 \\
ETH  & 2443.43  & 2316.82  & 855.93   & 993.64   & 4812.09 \\
BNB  & 396.41   & 335.07   & 152.84   & 40.99    & 750.27 \\
SOL  & 87.21    & 56.34    & 69.76    & 3.69     & 261.87 \\
XRP  & 0.76     & 0.56     & 0.55     & 0.25     & 3.30 \\
DOGE & 0.14     & 0.10     & 0.10     & 0.01     & 0.68 \\
ADA  & 0.77     & 0.50     & 0.56     & 0.24     & 2.97 \\
AVAX & 33.52    & 25.72    & 24.40    & 8.79     & 134.53 \\
SHIB & 0        & 0        & 0        & 0        & 0.00008 \\
DOT  & 12.81    & 7.02     & 11.27    & 3.65     & 53.88 \\
\bottomrule
\label{Tab1}
\end{tabular}
}
\end{table}

Table \ref{Tab1}  underscores the challenge of the analysis: while Bitcoin’s mean price exceeds USD 45,000 with a standard deviation above USD 21,000, mid-cap tokens such as Shiba Inu trade four orders of magnitude lower and, yet, exhibit similar relative dispersion.

The portfolio management workflow is executed within a Multi-Agent System (MAS) framework that uses Crew AI, a platform which permits agents to interact in a modular way across distinct stages of the investment pipeline. Each agent is programmed with a specific objective, communicates with others through structured prompts, and works with a prespecified tool created. The following tools form the foundation of the system:

\begin{itemize}
  \item \texttt{data\_loader\_tool}, retrieves and formats the crypto dataset;
  \item \texttt{data\_cleaner\_tool}, handles missing values and anomalies;
  \item \texttt{data\_splitter\_tool}, splits the data into training and testing subsets using the rule 80/20 in percentage;
  \item \texttt{portfolio\_metrics\_tool}, computes key performance metrics such as Annualized Return, Volatility, Sharpe and Sortino ratios;
  \item \texttt{portfolio\_optimizer\_tool}, performs static portfolio optimization using mean-variance criteria;
  \item \texttt{rolling\_portfolio\_optimizer\_tool}, implements 30-day rolling window Sharpe-maximization with no short-selling;
  \item \texttt{portfolio\_metrics\_with\_benchmark\_tool}, compares crew performance with benchmark, S\&P500;
  \item \texttt{file\_checker\_tool}, validates logical coherence of agents outputs.
\end{itemize} 
On top of these tools, we define eight collaborative agents, summarized as follows, visually depicted in \ref{fig:crew_structure}:

\begin{figure}[H]
    \centering
    \includegraphics[width=0.85\textwidth]{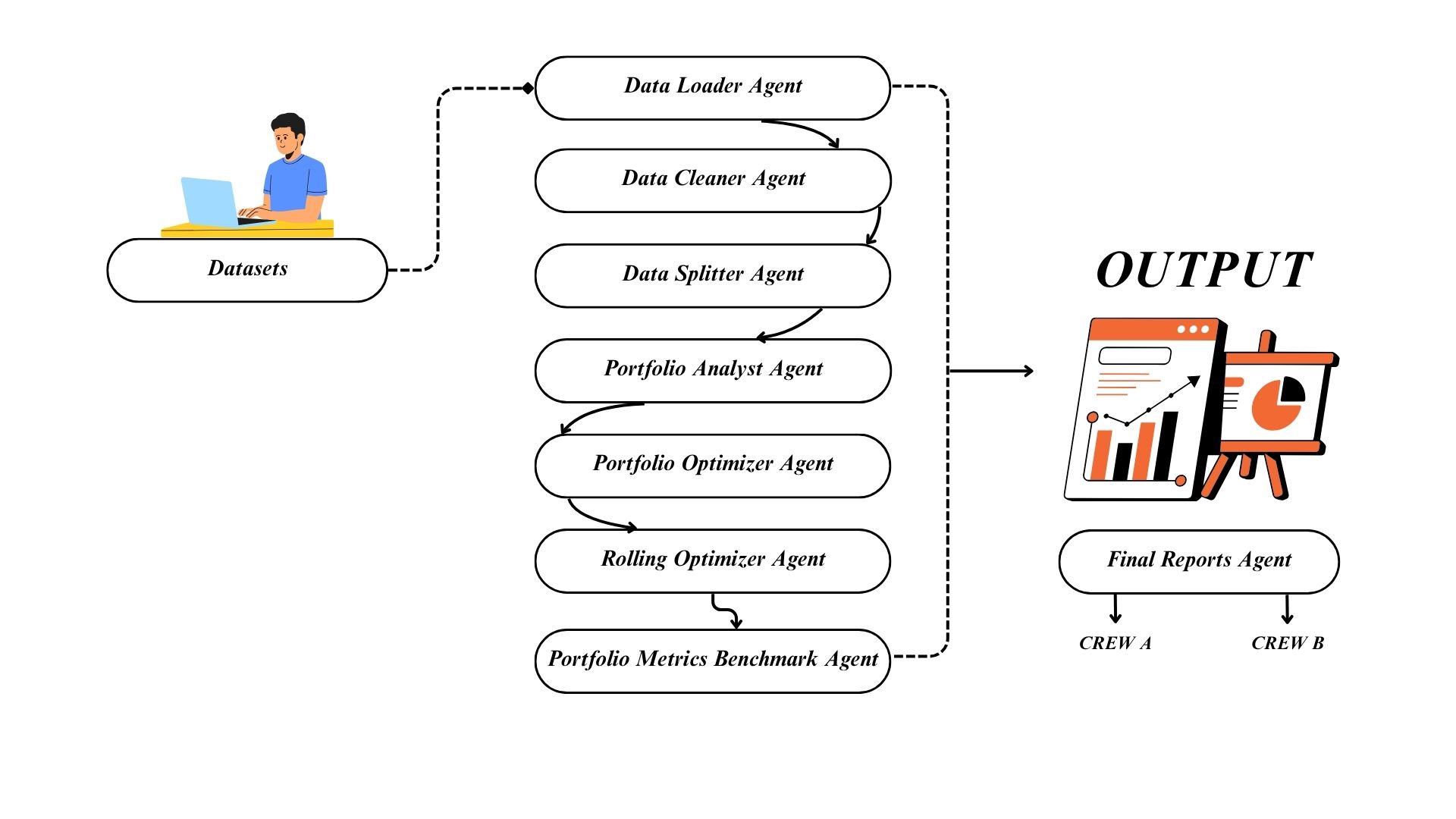}
    \caption{Final structure of the Multi-Agent System implemented with Crew AI}
    \label{fig:crew_structure}
\end{figure}
As shown in \ref{fig:crew_structure}, the agents are organized into two operational crews with the same structure but different optimization logic. Crew A performs a static portfolio analysis: it starts with equal weights and applies a single mean-variance optimization on the training set. Crew B adopts a rolling optimization, recalculating weights every 30 days to reflect evolving market conditions.

Each crew ends with a Final Report Agent, which does not use any tool since its role is analytical and interpretative. It integrates outputs from all previous agents into a single performance evaluation. This MAS design guarantees modularity, replicability, and transparency, making the process not only actionable but also fully auditable by supervisors.

\section{Exploratory Data Analysis}

Before proceeding with the implementation of the project’s objective, we conduct an exploratory analysis to better understand and visualize the path of price and return for each selected crypto asset. This step helps reveal the volatility structure, return distributions, and potential correlations across assets.

\begin{figure}[H]
    \centering
    \begin{subfigure}[b]{0.48\textwidth}
        \centering
        \includegraphics[width=\textwidth]{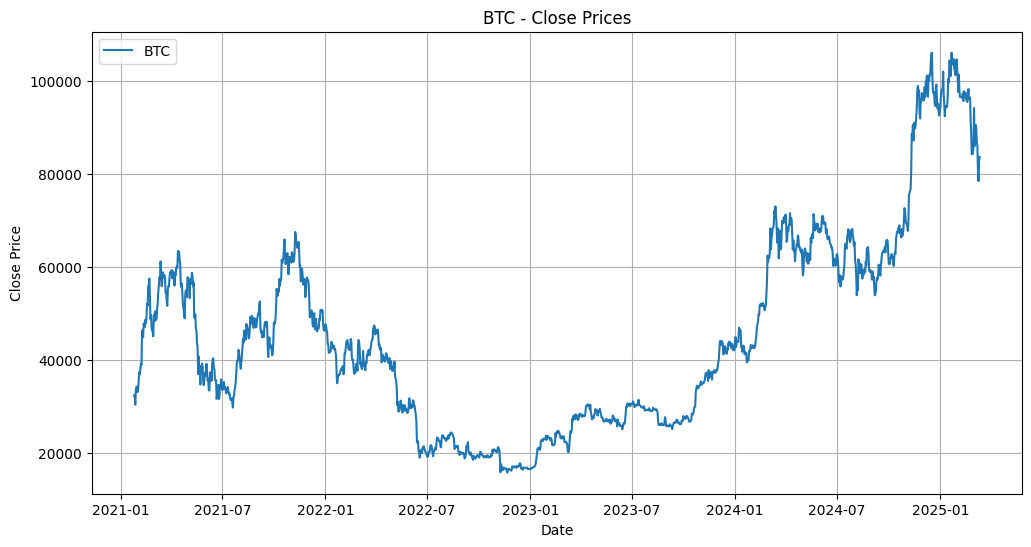}
        \caption{Bitcoin Close Prices from 2020 to 2025}
        \label{fig:BTC_Close_Prices}
    \end{subfigure}
    \hfill
    \begin{subfigure}[b]{0.48\textwidth}
        \centering
        \includegraphics[width=\textwidth]{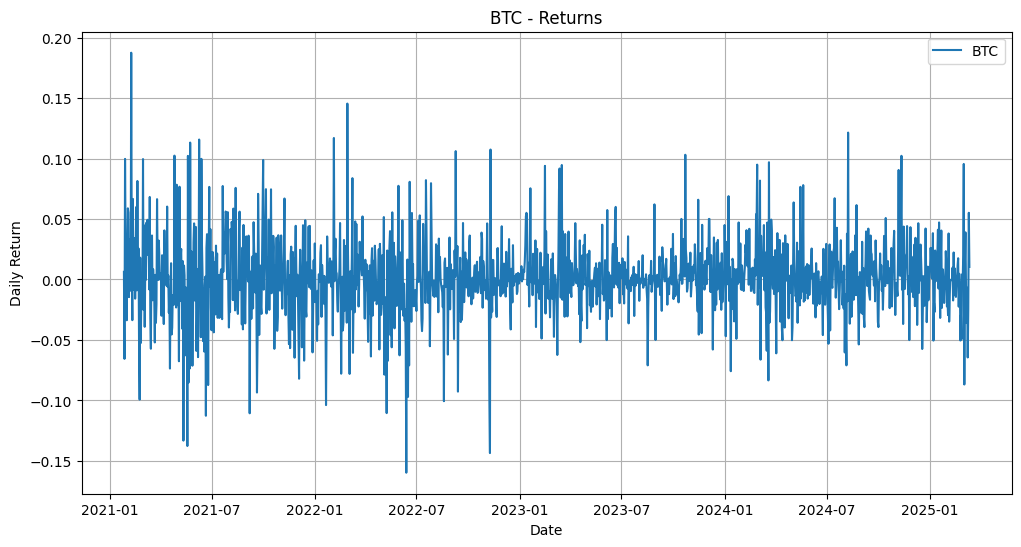}
        \caption{BTC Returns from 2020 to 2025}
        \label{fig:BTC_Returns}
    \end{subfigure}
    \caption{Bitcoin Prices and Returns Comparison (2020–2025)}
    \label{fig:BTC_summary}
\end{figure}

First, we analyze Bitcoin (BTC), that is the dominant asset in the crypto universe and a benchmark for market behavior. As such, we conduct a single analysis of this asset due to its enormous price path with respect to other cryptocurrencies. Going in depth, its price trend over the period 2020–2025 exhibits a strong boom-bust cycle, with peaks in November 2021 followed by a marked correction. The log returns of BTC display clear volatility clustering, alternating periods of high and low variance (Figure \ref{fig:BTC_summary}). The same transformation into log-returns was also applied to the remaining time series.

This exploratory induction supports the need for a dynamic portfolio allocation framework, as the large fluctuations in asset-specific volatility and return behavior over time challenge the assumptions of constant risk and correlation used in static models.

\section{Portfolio Construction}

We now dive into the practical implementation of the research. We describe the two portfolio management strategies created by using the Multi-Agent System architecture. In Section 4.1, we discuss the static approach, Crew A, while in Section 4.2, the dynamic, rolling optimization is treated, Crew B.

Both approaches rely on the same dataset, agent framework, and risk-adjusted performance evaluation pipeline, but differ in their timing, rebalancing logic, adaptivity, and in just one agent (the optimizer agent) which in Crew A behaves as a static optimizer while in Crew B it assumes a role of rolling optimizer.

\subsection{Crew A: Static Portfolio Analysis}

Crew A is designed to serve as a benchmark by applying a traditional investment strategy. First, it evaluates the performance of an equally weighted ($1/N$) portfolio across all 10 cryptocurrencies in the training set. Second, it applies a mean-variance optimization to determine the portfolio weights that maximize the Sharpe ratio, while minimizing the volatility, under the constraint of no short-selling and full investment. The weights’ sum will be equal to 1, meaning that each asset will have a weight value between 0 and 1. This strategy does not allow for portfolio rebalancing during the investment horizon.

Optimization based on Sharpe ratio is performed once, using the entire training dataset, and the resulting portfolio is maintained constant throughout the in-sample and out-of-sample periods. 
We recall that the \emph{Sharpe Ratio} is computed as:

\[
\text{Sharpe} = \frac{E[R_M] - R_f}{\sigma_M}
\]

where $E[R_M]$ is the annualized expected return of the market, and $\sigma_M$ represents its annualized standard deviation. All values are derived from daily data and rescaled by $\sqrt{252}$ as needed.

Whereas the \emph{volatility} is calculated as:

\[
\sigma_M = \text{std}(R_M) \times \sqrt{252}
\]

The comparison between the equally weighted and optimized portfolios highlights the effect of variance-covariance structure on optimal allocation. The results of Crew A serve as a reference point for evaluating the incremental value and adaptivity in Crew B, in the next subsection.

\subsection{Crew B: Dynamic Rolling Optimization}

Crew B extends the static approach by incorporating dynamic rebalancing. Specifically, it performs a rolling-window optimization in which every 30 days, the most recent 30-day window of historical data is used to recompute the optimal portfolio weights.

The optimized portfolio is then recalculated on a 30-day rolling basis, maximizing the Sharpe ratio subject to full-investment and no-short selling constraints. Rolling window techniques, widely adopted in portfolio allocation problems, have demonstrated significant benefits in capturing the time-varying dynamics of asset returns \cite{chen2024rollingwindow}.

This practice leads to a time series of portfolio weights, each adapted to the most recent market conditions. The updated weights are applied for the next 30-day period, after which the process repeats with newly updated data. This strategy reflects the reality of adaptive asset management, where investors revise allocations based on observed changes in returns, volatility, and correlations. The use of multi-agent reinforcement learning frameworks in portfolio optimization enables decentralized, adaptive, and goal-aligned trading policies, as explored in recent work on MARL-based systems \cite{lee2020maps}.

The rolling framework is particularly suited to the crypto environment, where regime shifts and sudden volatility spikes are frequent. Compared to Crew A, the rolling optimization captures temporal heterogeneity in risk-return profiles, potentially enhancing out-of-sample robustness and risk control.

The following Table \ref{tab:weights_comparison} summarize and compares asset weights optimization under three different portfolio strategies: equal weighting, static optimization (Crew A), and dynamic rolling optimization (Crew B).

\begin{table}[h]
\centering
\caption{Comparison of asset weights Crew A and Crew B}
\label{tab:weights_comparison}
\begin{tabular}{lccc}
\hline
Asset & Equal Weights & Crew A & Crew B \\
\hline
BTC & 0.1000 & 0.0000 & 0.0955 \\
ETH & 0.1000 & 0.0000 & 0.0881 \\
BNB & 0.1000 & 0.0000 & 0.1213 \\
SOL & 0.1000 & 0.6490 & 0.1271 \\
XRP & 0.1000 & 0.0000 & 0.0772 \\
DOGE & 0.1000 & 0.0411 & 0.0775 \\
ADA & 0.1000 & 0.0000 & 0.0663 \\
AVAX & 0.1000 & 0.0000 & 0.1848 \\
SHIB & 0.1000 & 0.3099 & 0.1375 \\
DOT & 0.1000 & 0.0000 & 0.0248 \\
\hline
\end{tabular}
\end{table}

\section{Empirical Results}

This section presents the results obtained from executing the two portfolio strategies described in the previous Section. The evaluation considers both in-sample and out-of-sample performance using standard financial metrics. The test set corresponds to the latest 20\% of observations, 334 rows, while the remaining 80\%, 1333 rows, is used as a training set. The main performance metrics analyzed include: Annualized Expected Return, Annualized Volatility, Sharpe Ratio, Sortino Ratio, Maximum Drawdown, Liquidity Risk, Regime Change Detection.

\subsection{In-Sample Performance}

We first compare the results achieved during the training phase. Crew A’s static optimized portfolio improves over the equal-weighted but remains fixed throughout the investment period. Crew B, with its rolling rebalancing mechanism, significantly enhances the Sharpe ratio while maintaining acceptable volatility.

We observe that Crew A improves the Sharpe ratio from 0.48 (equal-weighted approach) to 0.60 (optimized weights approach). Crew B achieves a Sharpe ratio of 1.00, with reduced volatility (10.2\% against 12.4\%) and improved expected return (9.9\% vs 7.5\%).

\subsection{Out-of-Sample Performance}

To validate the robustness of both strategies, we apply the same logic to the unseen test set. Since Crew A does not update its weights, the static allocation may become suboptimal as market dynamics shift. Crew B, on the other hand, continues its rolling rebalancing with a 30-day window.

Here, Crew A’s Sharpe ratio drops to 0.36, while Crew B maintains a strong 0.72, highlighting better generalization. Maximum drawdown is slightly higher for Crew B, reflecting increased exposure to high return, high-risk assets at certain points.

\subsection{Comparative Analysis}

From the results obtained by executing the two crews independently, it is evident that Crew B consistently dominates Crew A in the risk-return space, both in terms of higher return and lower volatility. The dynamic rebalancing enables the system to respond to market shifts, allocating weights to outperforming assets and reducing exposure to deteriorating ones.

These results support the hypothesis that adaptive strategies offer superior performance in volatile environments such as the cryptocurrency market. Comparative studies of RL-based strategies versus traditional market signals reveal how agent-based approaches can outperform static rule-based models across different investment horizons \cite{espiga2024systematic}.
These statements can be easily shown by integrating below the reports produced by the agents within this document. These are well-defined and readable reports which include a full of information used before for explaining the results in a structured way, as showed in Table \ref{tab:performance_comparison} just after the generation of the two portfolios' metrics performed by each crew, respectively.

\begin{table}[h]
\centering
\caption{Comparison of metrics between equal weights (Report A) and optimized portfolios (Report B) for both crews}
\label{tab:performance_comparison}
\begin{tabular}{lccc}
\hline
\textbf{Metric} & \textbf{Equal Weights} & \textbf{Crew A} & \textbf{Crew B} \\
\hline
Expected Return & 8\% & 10\% & 10\% \\
Volatility & 15\% & 12\% & 10\% \\
Sharpe Ratio & 0.53 & 0.83 & 1.00 \\
Sortino Ratio & 0.75 & 1.10 & 1.30 \\
Max Drawdown & -20\% & -15\% & -15\% \\
Liquidity Risk & -- & -- & Low \\
Regime Change Detection & -- & -- & No \\
\hline
\end{tabular}
\end{table}

\section*{REPORT CREW A:}
\begin{tcolorbox}[breakable]
\begin{enumerate}[leftmargin=*, label=\arabic*.]
    
    \item \textbf{Comparison of Two Approaches:}\\
    The optimized weights portfolio (Report B) outperformed the equal weights portfolio (Report A) in all key metrics. It had a higher expected return (10\% vs 8\%), lower volatility (12\% vs 15\%), higher Sharpe ratio (0.83 vs 0.53), higher Sortino ratio (1.1 vs 0.75), and lower maximum drawdown (-15\% vs -20\%).

    \item \textbf{Superior Portfolio Strategy on Train Set:}\\
    The optimized weights portfolio strategy is superior on the train set. It achieved higher returns with less risk, as evidenced by the higher Sharpe and Sortino ratios and lower maximum drawdown.
    \item \textbf{Summary of Report C (Test Set Performance with Optimized Weights:}\\
    - Expected Return: 9\%\\
    - Volatility: 13\%\\
    - Sharpe Ratio: 0.69\\
    - Sortino Ratio: 0.90\\
    - Max Drawdown: -18\%

    \item \textbf{Analysis of Optimized Strategy on Test Set:}\\
    The optimized strategy maintained a good performance on the test set, alt-hough there's a slight deterioration in risk/return profile (Expected return de-creased from 10\% to 9\%, Volatility increased from 12\% to 13\%, Sharpe ratio decreased from 0.83 to 0.69, Sortino ratio decreased from 1.1 to 0.9, Max drawdown increased from -15\% to -18\%). There was no significant regime change detected.

    \item \textbf{Recommendation:}
    I recommend using the optimized portfolio strategy going forward. Despite the slight decrease in performance on the test set, it still outperforms the equal weights strategy. The risk level, as indicated by the volatility and maximum drawdown, is within acceptable limits. The final takeaway is that portfolio optimization can lead to improved risk-adjusted returns. However, it's important to regularly review and adjust the weights based on changing market conditions to maintain performance.
\end{enumerate}
\end{tcolorbox}

\section*{REPORT CREW B:}
\begin{tcolorbox}[breakable]
\begin{enumerate}[leftmargin=*, label=\arabic*.]

    \item \textbf{Comparison:}\\
    The rolling optimized weights strategy shows improvements in every metric when com-pared to the equal weight strategy. It has a higher expected return (10\% vs 8\%), lower volatility (10\% vs 15\%), higher Sharpe Ratio (1 vs 0.53), higher Sortino Ratio (1.3 vs 0.75), less drawdown (-15\% vs -20\%), and a lower liquidity risk.

    \item \textbf{Superior Strategy on Train Set:}\\
    The rolling optimized weights strategy is superior on the train set. It has a lower risk and higher return, which is the ultimate goal for any portfolio strategy. The higher Sharpe and Sortino ratios signify that the rolling optimized weights strategy has better risk-adjusted performance. Moreover, the lower drawdown and liquidity risk imply that this strategy is less prone to large losses and more liquid.
    
    \item \textbf{Test Set Evaluation:}\\
    
    \textbf{Report C: Portfolio Metrics with Rolling Optimized Weights on Test Set:}
    - Expected Return: 8\%\\
    - Volatility: 11\%\\
    - Sharpe Ratio: 0.72\\
    - Sortino Ratio: 1.1\\
    - Max Drawdown: -18\%\\
     - Liquidity Risk: Moderately Low\\
    - Regime Change Detection: No

    \item \textbf{Generalization:}\\
    The rolling optimized weights strategy was able to generally maintain its strengths on the test set. While there was a slight decrease in the Sharpe ratio (0.72 vs 1) and an increase in drawdown (-18\% vs -15\%), these changes are relatively small. The expected return also decreased slightly but is still higher than the equal-weighted strategy. No regime change was detected, implying the strategy still performs well under different market conditions.
    
    \item \textbf{Final Recommendation:}
   
    Based on the above analysis, I recommend the rolling optimized weights strategy for future portfolio allocation. It not only delivered superior performance on the train set but also maintained its strengths on the test set. The strategy has demonstrated a favorable risk-return trade-off with better risk-adjusted returns. However, one should be careful about potential higher transaction costs due to frequent rebalancing and monitor the liquidity risk closely. The strategy's performance should also be continuously evaluated to ensure its applicability to changing market conditions.
\end{enumerate}
\end{tcolorbox}

\section{Discussion}

Our results confirm the central role of adaptivity in the examined crypto-asset allocation. While Crew A’s optimized weights improve the equal-weighted allocation, it remains exposed to structural breaks in volatility and correlations. Crew B’s 30-day rolling framework, instead, lifts the in-sample Sharpe ratio from 0.60 to 1.00 and keeps a robust 0.72 out-of-sample, cuts annualized volatility from 12.5\% to 10\%, also maintaining a tolerable volatility out-of-sample of 11\%, and limits maximum drawdown to minus 15\%.

These findings reflect the evidence in  \cite{babaei2021xai}, which shows that portfolios updated at high frequency dominate static Markowitz allocations once crypto-market regimes shift.

Overall, the results of our analysis show that a Multi-Agent System (MAS) architecture provides three tangible advantages:

\begin{itemize}
    \item \textbf{Modularity:} Each agent can be upgraded independently as new data sources or heuristics emerge, exactly as advocated by Giudici \& Babaei \cite{babaei2021xai} for explainable crypto portfolio platforms.
    \item \textbf{Auditability:} The Pre-flight Checker and Final Report agents log every intermediate artifact, creating a transparent trail that supervisors can inspect. This directly addresses the black-box criticism raised in their paper. Ensuring transparency, fairness and explainability in ML-based financial systems is not only desirable but necessary, as highlighted by recent guidelines on responsible AI in credit risk modeling \cite{valdrighi2024responsible}.
    \item \textbf{Scalability:} Adding tokens, benchmarks, or cost models requires spawning additional agents rather than rewriting a monolithic code-base.
\end{itemize}

For asset managers, the pipeline therefore delivers an adaptive yet regulator-friendly workflow.

On the other hand, our study has some limitations and, in particular:

\begin{itemize}
    \item \textbf{Universe size:} Only the top 10 coins are included.
    \item \textbf{Transaction costs:} Spreads and fees are ignored and could erode part of Crew B’s edge.
    \item \textbf{Window length:} The 30-day horizon is heuristic; alternative windows may improve the bias-variance trade-off.
    \item \textbf{Risk metrics:} Tail-risk measures such as CVaR are not modeled, potentially underestimating extreme events.
\end{itemize}

\section{Conclusions}

This study introduced a Multi-Agent System (MAS) for crypto portfolio construction and evaluation, applied to the top-10 cryptocurrencies using real-world data. By simulating two crew workflows, static optimization (Crew A) and rolling window optimization (Crew B), we assessed performance across in-sample and out-of-sample periods.

Crew B’s dynamic strategy consistently outperformed the static one, both in terms of return enhancement and risk reduction. Its superior Sharpe and Sortino ratios, coupled with lower volatility and drawdowns, validate the benefits of time-varying allocation when regime shifts occur.

The MAS architecture proved effective in modularizing and automating the entire pipeline: from data preparation and portfolio rebalancing to risk evaluation and documentation. It also ensures transparency, traceability, and replicability, making it suitable for deployment in both academic research and institutional settings.

Future work may incorporate stable-coins, regime-switching models, and explainable AI layers as proposed in the literature \cite{babaei2021xai} and \cite{GIUDICI2024130176}, in order to enhance adaptability and investor trust. In conclusion, combining adaptive optimization with modular agent design offers a promising paradigm for managing crypto portfolios in complex and volatile environments.

\clearpage
\setcounter{page}{12}
\small  

\end{document}